\documentclass[letterpaper, 10 pt, conference]{ieeeconf}  

\IEEEoverridecommandlockouts                              
\usepackage{cite}
\usepackage{amsmath,amssymb,amsfonts}
\usepackage[ruled,vlined]{algorithm2e}
\usepackage{graphicx}
\usepackage{textcomp}
\usepackage{xcolor}
\usepackage{gensymb}
\usepackage{subcaption}
\usepackage{array}
\usepackage{romannum}
\usepackage{algpseudocode}
\usepackage{amsmath}
\usepackage{float}
\usepackage{booktabs}
  \usepackage{amsthm}
\newtheorem{remark}{Remark}

\title{\LARGE \bf
Towards Input-Convex Neural Network Modeling for Battery Optimization in Power Systems
}

\author{Arash Omidi, Tanmay Mishra, and Mads R. Almassalkhi
\thanks{This material is based upon work supported by the U.S. Department of Energy’s Office of EERE under award number DE-EE0010147. The views expressed herein do not necessarily represent the views of the U.S. Department of Energy or the United States Government.}
\thanks{Arash Omidi, Tanmay Mishra, and Mads R. Almassalkhi are with the Department of Electrical and Biomedical Engineering, 
        University of Vermont, Burlington, Vermont, USA
        {\tt\small \{aomidi,tmishra,malmassa\}@uvm.edu}}%
}

\begin{document}

\maketitle
\thispagestyle{empty}
\pagestyle{empty}

\begin{abstract}
Battery energy storage systems (BESS) play an increasingly vital role in integrating renewable generation into power grids due to their ability to dynamically balance supply. Grid-tied batteries typically employ power converters, where part-load efficiencies vary non-linearly. While this non-linearity can be modeled with high accuracy, it poses challenges for optimization, particularly in ensuring computational tractability. In this paper, we consider a non-linear BESS formulation based on the Energy Reservoir Model (ERM). A data-driven approach is introduced with the input-convex neural network (ICNN) to approximate the nonlinear efficiency with a convex function. The epigraph of the convex function is used to engender a convex program for battery ERM optimization. This relaxed ICNN method is applied to a battery revenue maximization problem and is compared with three other ERM formulations (nonlinear, linear and mixed-integer). Specifically, ICNN-based method appear to be promising for future battery optimization with desirable feasibility and optimality outcomes across revenue maximization use-case.
\end{abstract}

\section{INTRODUCTION}
As global efforts to decarbonize intensify in response to climate change, the integration of renewable energy sources has become critical. However, the intermittent nature of renewables such as solar and wind presents significant challenges to maintaining grid stability and reliability. Battery energy storage systems (BESS) have emerged as a pivotal solution, offering numerous advantages such as rapid response time, high energy density, and efficiency. They provide grid operators with a wide range of services, including renewable energy smoothing, frequency regulation, voltage support, and black-start capability \cite{IEA}. Maximizing the value of these services, however, requires the development of accurate BESS models that account for state of charge (SoC) dynamics, power, voltage, and temperature. Accurate modeling of BESS considering these factors presents considerable challenges \cite{Byrne}.

Various BESS models have been proposed to represent the physical processes of electrical energy conversion and storage~\cite{Rosewater}. These models can be classified into four main categories: i) Energy Reservoir Models (ERMs), ii) Charge Reservoir Models (CRMs) or Equivalent Circuit Models (ECMs), iii) Electrochemical-based models, and iv) Data-driven models. The ERM treats the battery and inverter as an energy reservoir focusing on the dynamic relationship between SoC and charging and discharging power without explicitly considering internal electrochemical dynamics. While ECMs use passive electrical components such as resistors and capacitors to model the battery, they require frequent parameter adjustments due to variability in operating conditions. Electrochemical-based models are high-order representations of internal physical processes (e.g., PDEs, SDEs) that place significant burdens on computing and require high-fidelity battery (cell) data~\cite{Rosewater}.  Data-driven models, utilizing machine learning techniques, accurately predict the battery’s nonlinear behavior by training on extensive datasets~\cite{Lucaferri}. 

In power systems, the ERM is often employed to model the combined battery and inverter system performance to drive the SoC trajectories based on controlled charging and discharging inputs (powers), while accounting for energy losses due to inverter efficiencies. In many cases, the ERM is engineered to assume no losses or constant efficiency, leading to model mismatches~\cite{Huang,Nazir,mazenCDC2023}. These modeling decisions are effective in power system planning~\cite{Shin, Siano, Emrani}, but are less suitable for power system operations and control where non-linearities due to inverter switching losses and internal resistance, and temperature variations affect battery SoC evolutions~\cite{King,Aaslid2020}. Additionally, there are considerations around physical limitations on simultaneous charging and discharging that give rise to non-linear complementarity constraints~\cite{mazenCDC2023}.

Incorporating these nonlinear efficiency relations in the ERM introduces non-convexity in power system optimization. This issue is addressed by different relaxation methods in the literature. In~\cite{Do}, a piecewise linear efficiency model was proposed for the unit commitment problem. The results demonstrate that accounting for nonlinear efficiencies can significantly reduce errors between planned and actual operational costs relative to constant-efficiency models. 
A dynamic programming method was proposed to address the nonlinearity and non-convexity of efficiencies in a revenue maximization problem~\cite{Nguyen}. This method breaks the problem into smaller sub-problems and employs a forward-search strategy to iteratively optimize the SoC trajectory. However, as the scale of the problem increases (e.g., longer time horizon, more batteries), a significant computational burden arises. Approximating the nonlinear efficiencies by polynomial approximations have been employed in predictive control and optimization settings~\cite{Morstyn,Aaslid2020}. These approximations can embody different problem structures (e.g., convex or non-convex) depending on their design and the data. 

Given the challenges related to accurately modeling the nonlinear and dynamic behaviors of BESS due to complex physical phenomena and operational constraints, there is a need for modeling approaches that can capture these complexities without imposing significant computational burdens. Data-driven approaches for battery modeling leverage machine learning techniques to capture complex, nonlinear behaviors from large datasets. Compared to physics-based models, these methods provide greater flexibility and adaptability~\cite{Lucaferri}. Integrating data-driven models into optimization problems can enhance battery performance and extend its life-cycle. Specifically, input-convex neural networks (ICNNs) are designed so that the function mapping input to output is convex. This makes them particularly suitable for optimization applications~\cite{Amos}. ICNNs can directly learn from data, enabling the approximation of highly complex relationships without requiring predefined functional forms. They have been effectively applied to various optimization problems in power systems and control~\cite{Chen, Dvorkin}. In this paper, we model the nonlinear and non-convex behavior of efficiencies of BESS using ICNNs. This study emphasizes the importance of accurately capturing the nonlinear behavior of BESS efficiencies, particularly in short-term and real-time control applications.
The main contributions of this paper are:
 \begin{itemize}
    \item propose a novel ICNN-based BESS formulation to characterize the underlying nonlinear ERM model. 

    \item engineer the ICNN-based formulation within a tractable optimization formulation.

    \item validate the proposed ICNN-based BESS models through comprehensive simulation-based analysis in a practical use-case, demonstrating superior performance in terms of computational complexity and optimality.
    
\end{itemize}
The remainder of this paper is organized as follows: Section II discusses two representations of battery model: Nonlinear ERM and ICNN approximation of ERM. Section III outlines four formulations for battery dispatch optimization: nonlinear ERM, linear ERM, relaxed ICNN, and Big-M formulation for ICNN. Section IV presents the revenue maximization use-case, which is employed to compare these four formulations. In this section, the simulation results, along with a discussion, are presented. Finally, Section V offers conclusions and suggestions for future work.


\section{Battery Energy Storage Modeling}
\begin{figure}[!tbp]
    \centering
    \includegraphics[width=0.45\textwidth]{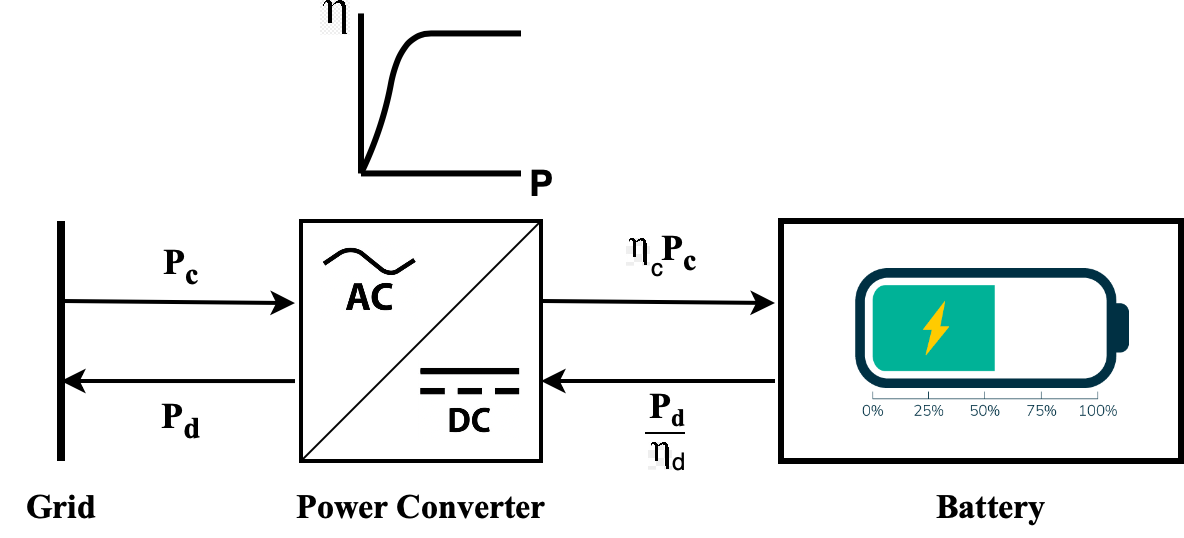}
    \caption{The block diagram of BESS}
    \label{BESS}
    \centering
\end{figure}

The system of interest in this paper is an AC grid-tied battery and inverter as shown in Fig.~\ref{BESS}. The AC power imported from the grid during charging is $P_{\text{c}}$, while the AC power exported during discharging is $P_{\text{d}}$. Combined energy losses of the power converter and the battery are captured by the charging and discharging efficiencies  $\eta_{\text{c}}, \eta_{\text{d}} \in (0,1]$, respectively. In this section, two BESS models are discussed: a) non-linear ERM, and b) a data-driven ICNN model. The nonlinear ERM estimates the SoC by considering the charging and discharging power with non-linear efficiencies, while the ICNN model treats the nonlinear ERM relations with a convex approximation.


\subsection{Energy Reservoir Model (ERM)}
The ERM 
is widely used for analyzing battery operation in techno-economic studies in power systems~\cite{Vykhodtsev}. The SoC dynamics are defined by
\begin{equation}
    E(t) = E(t_\text{0}) + \frac{1}{C_{\text{batt}}} \int_{t_\text{0}}^{t} (\eta_\text{c}(\tau)  P_\text{c}(\tau) - \frac{P_\text{d}(\tau)}{\eta_\text{d}(\tau)}) \, d\tau
    \label{Soc_1}
\end{equation}
where $E(t) \in [\underline{E},\overline{E}]$ is the SoC at time $t$, $E(t_0)$ denotes initial conditions, $P_\text{c}(\tau), P_\text{d}(\tau) \in [0,\overline{P}]$ are the charging and discharging power at time $\tau$, respectively. 
The charging and discharging efficiencies at time $\tau$ are given by $\eta_\text{c}(\tau), \eta_\text{d}(\tau) \in (0,1]$, respectively, and $C_{\text{batt}}$ is the nominal capacity of the battery in kWh. 

For our optimization framework, the continuous-time model in~\eqref{Soc_1} is discretized using zero-order hold sampling with time step $\Delta t$. For example, $P_{\text{c,d}}(t) = P_{\text{c,d}}[k]  \quad \forall t \in [k \Delta t , (k+1) \Delta t)$. The discretized SoC equation is given by:
\begin{equation}
    E[k+1] = E_\text{0} + \frac{\Delta t}{C_{\text{batt}}} \sum_{l=0}^{k}(\eta_\text{c}[l] P_\text{c}[l] - \frac{P_\text{d}[l]}{\eta_\text{d}[l]}  ).
    \label{Soc_3}
\end{equation}

\begin{figure}[!tbp]
\centering
\includegraphics[width=0.45\textwidth]{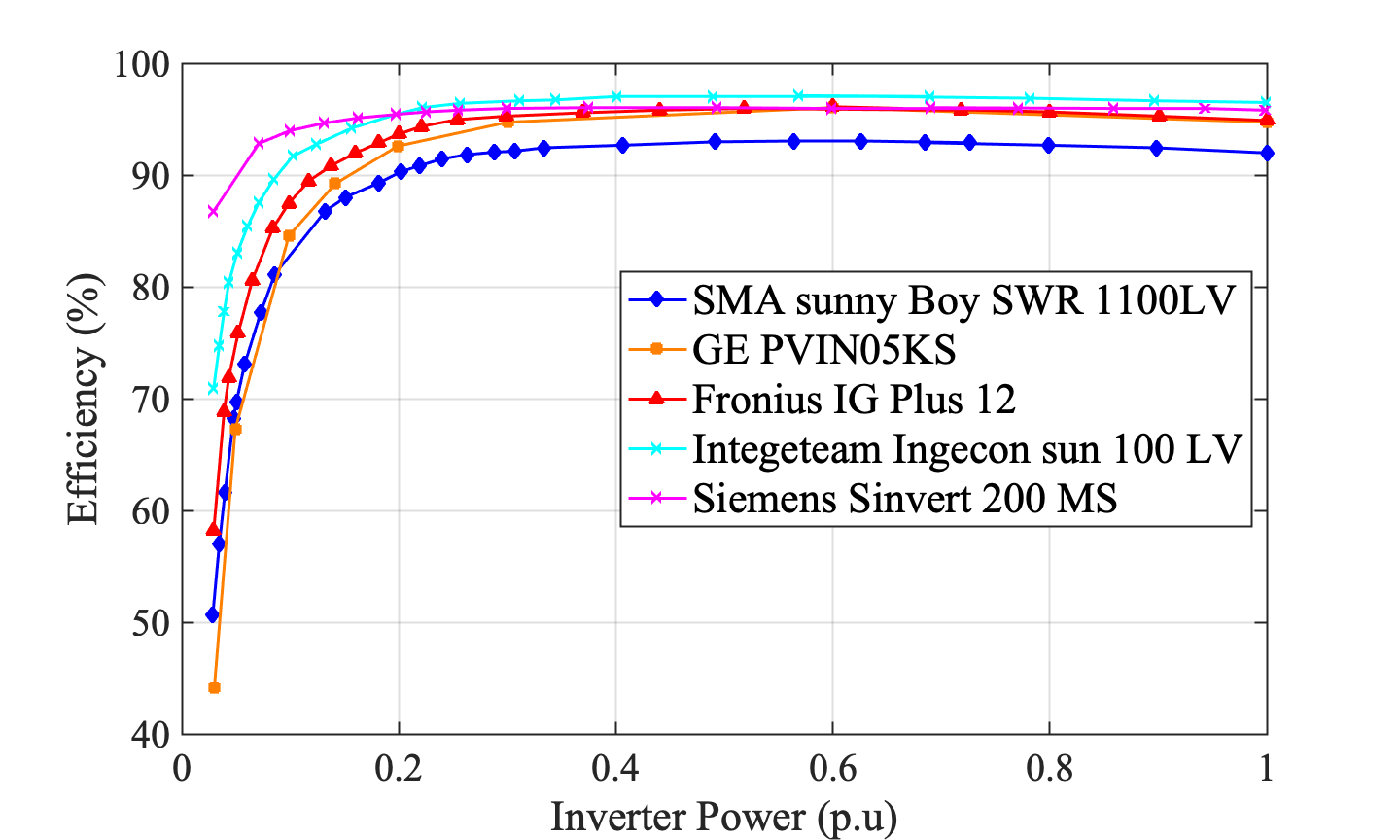}
\caption{Efficiency curve for commercial inverters \cite{Faranda}}
\label{Eff}
\centering
\end{figure}

The efficiencies, $\eta_\text{c}[l]$ and $\eta_\text{d}[l]$ in~\eqref{Soc_3}, are often assumed constant. But as seen in Fig.~\ref{Eff}, it is evident that the efficiency is nonlinear~\cite{Faranda}. Specifically, at low part-load operation ($<0.3$ p.u.), the efficiencies enter the non-linear operating region. Thus, we denote the efficiencies as a function of the grid exchange with the general functions
\begin{align}
    & f_\text{c}(P_\text{c}[l]) :=\eta_\text{c}(P_\text{c}[l]) P_\text{c}[l] \\
    &g_\text{d}(P_\text{d}[l]) :=\frac{P_\text{d}[l]}{\eta_\text{d}(P_\text{d}[l])}.
\end{align}

To model these part-load relations, $f_{\text{c}}(.)$ and $g_{\text{d}}(.)$, various approaches have been developed, including constant-value approximation~\cite{Shin}, linear fit~\cite{Gonzalez}, piece-wise linear (PWL) approximation~\cite{Do}, and the full nonlinear relations as either quadratic polynomials~\cite{Rosewater-eff} or exponential functions~\cite{Tanmay}. While simpler models may offer computational convenience, nonlinear functions more accurately capture the relations~\cite{Nguyen}. However, within a battery optimization setting, the nonlinearity begets a non-convex problem, which can be NP-hard.

Various methods, such as linearizations~\cite{Vaska} and convex relaxations~\cite{Abelova}, have been commonly proposed to address non-convexity in optimization problems related to efficiencies.  However, most of these approaches tend to oversimplify the nonlinear relationships, leading to inaccuracies in the modeling process, and increase the computational burden. To trade off on model complexity and accuracy, we consider a recently-developed method for the battery-inverter system that embeds a convex mathematical structure with data-driven (neural network) methods to create an input-convex neural network (or ICNN)~\cite{Amos}. 




\subsection{Input Convex Neural Network}

 Neural networks are extensively utilized in machine learning for their ability to approximate complex, nonlinear relationships. By structuring layers of neurons and applying parameters and nonlinear activation functions, neural networks are able to model various types of relationships. The ICNN is a 
 neural network architecture with constraints on the neural network’s weights to ensure that the output is always a convex function of the input~\cite{Amos}. This property makes ICNNs suitable for convex optimization problems. ICNNs have been effectively used in learning convex functions from data and integrating them into optimization frameworks in power system optimization \cite{Chen,Dvorkin}. 
 
Mathematically, an $N$-layer ICNN architecture, which defines a neural network over the input $x$, is defined by,
\begin{equation}
    z_{i} = g_i ( W_i z_\text{i-1} + D_i x + b_i)  ,  f(x;\theta) = z_\text{N},
\end{equation}
where $z_i$ represents the activation at layer $i=1,\hdots, N$ (with $z_0 = x$), and $\theta = \{W_{1:\text{N}}, D_{2:\text{N}}, b_{1:\text{N}}\}$ refers to the set of parameters (with $D_1 = 0$). The functions $g_i$ are the nonlinear activation functions used in the network. The function $f$ is guaranteed to be a convex function in $x$, provided that all $W_{2:\text{N}}$ are non-negative and all activation functions $g_i$ are convex and non-decreasing.
\begin{figure}[!t]
    \centering
\includegraphics[width=0.48\textwidth]{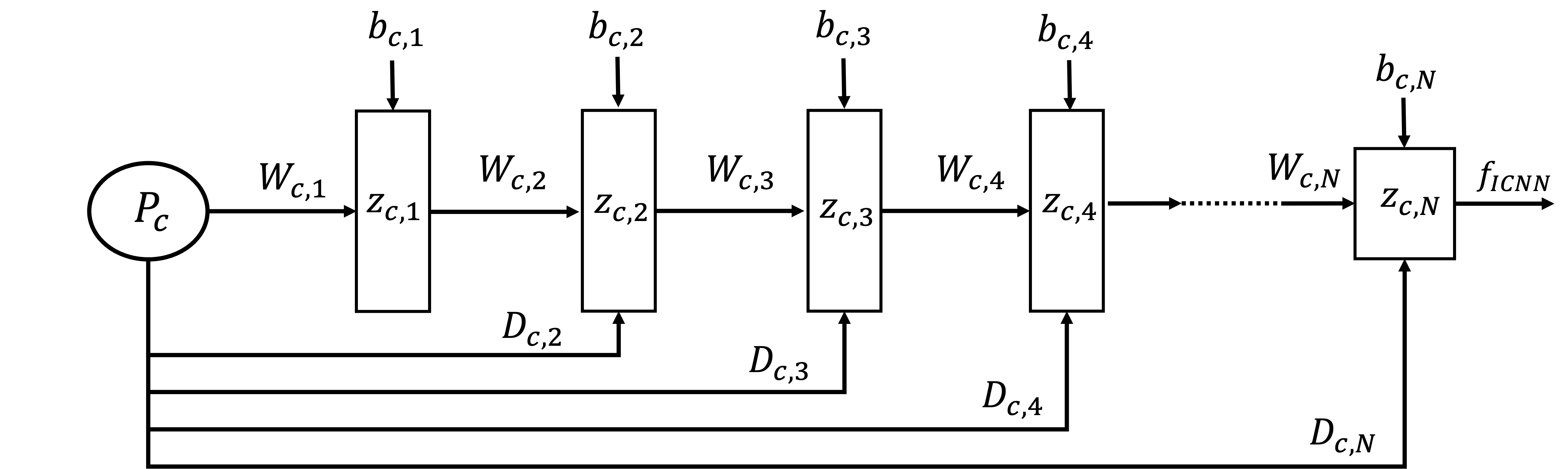}
    \caption{Block diagram of ICNN architecture for $f_\text{ICNN}$}
    \label{ICNN}
    \centering
\end{figure}

 The convexity of the proposed neural network is grounded in the composition rule for convex functions. This rule asserts that the composition of an inner convex function with another outer convex, non-decreasing function preserves convexity \cite{boyd}. A common example of such a function is the rectified linear unit (ReLU), which is both convex and non-decreasing. In this architecture, $W_{\text{2:N}}$ terms are constrained to be non-negative, while the $D_i$ terms can take negative values to offset any representational limitations imposed by the non-negativity of $W_{\text{2:N}}\geq 0$. Each layer $i$, combined with its bias term $b_i$, passes its output through the ReLU activation function before being forwarded to the next layer $i+1$.
 
To address the non-convexity in \eqref{Soc_3}, 
we consider each charging and discharging component separately. Consequently, the nonlinear portion can be approximated by the sum $f_{\text{ICNN}}(P_\text{c}) + g_{\text{ICNN}}(P_\text{d})$, representing two ICNN models for charging and discharging, respectively. Therefore, the $N$-layer ICNN model using ReLU activation functions is given by,
\normalsize 
 \begin{equation}
    E[k+1] = E[k] + \frac{\Delta t}{C_{\text{batt}}}(f_{\text{ICNN}}(P_\text{c}[k]) + g_{\text{ICNN}}(P_\text{d}[k])),
\end{equation}
where convex $f_{\text{ICNN}}(P_\text{c}[k])$ is given by
\begin{equation}
  f_{\text{ICNN}}[k] := W_{\text{c},N} z_{\text{c},N-1}[k] + D_{\text{c},N} P_\text{c}[k]+ b_{\text{c},N}, \label{f-ICNN}
\end{equation}
\noindent where $z_{\text{c},i}[k] = \text{max}(0,W_{\text{c},i} z_{\text{c},i-1} [k] +D_{\text{c},i} P_\text{c}[k]+b_{\text{c},i})$ for $ i = 1,...,N-1 $ (with $z_{\text{c},0}[k] = P_\text{c}[k]$ and $D_{\text{c},1} = 0$) and $W_{\text{c},i}$, $D_{\text{c},i}$, and $b_{\text{c},i}$ are the weights and biases of the layer $i$ of the neural network for charging part. The ICNN architecture for $f_{\text{ICNN}}(P_\text{c})$ is shown in Fig.~\ref{ICNN}. Similarly, convex $g_{\text{ICNN}}(P_\text{d}[k])$ is given by,
\begin{equation}
g_{\text{ICNN}}[k] = W_{\text{d},N} z_{\text{d},N-1}[k] + D_{\text{d},N} P_\text{d}[k]+ b_{\text{d},N}, \label{g-ICNN}
\end{equation}
where $z_{\text{d},i}[k] = \text{max}(0,W_{\text{d},i} z_{\text{d},i-1} [k] +D_{\text{d},i} P_\text{d}[k]+b_{\text{d},i})$ for $ i = 1,...,N-1 $ (with $z_{\text{d},0}[k] = P_\text{d}[k]$ and $D_{\text{d},1} = 0$) and $W_{\text{d},i}$, $D_{\text{d},i}$, and $b_{\text{d},i}$ are the weights and biases of the network layers for discharging part. 

Next, we embed this ICNN-based ERM in an battery optimization problem.

\normalsize
\section{Methodology}
\label{methods}
This section presents four battery optimization problem formulations: (i) Full NLP formulation, (ii) Constant-efficiency linear formulation, (iii) Relaxed ICNN formulation, and (iv) Big-M ICNN formulation. In these formulations, we let $f_\text{0}(P_\text{c}[k], P_\text{d}[k])$ denote an objective function whose specific form depends on the application.

\subsection{Full Nonlinear Formulation}
In this section, the full non-convex non-linear program (NLP) for the BESS is presented and is given by
\begin{subequations}
\begin{align}
\underset{P_\text{c}[k], P_\text{d}[k]}{\text{minimize}}&  \qquad f_\text{0}(P_\text{c}[k], P_\text{d}[k])\\
\text{subject to} & \notag \\
 E[k+1] &= E[k] + \frac{\Delta t}{C_{\text{batt}}} \left(f_\text{c}(P_\text{c}[k]) -  g_\text{d}(P_\text{d}[k]) \right)  \label{cons1} \\
  f_\text{c}(P_\text{c}&[k]) :=\eta_\text{c}(P_\text{c}[k]) P_\text{c}[k] \label{cons2}\\
    g_\text{d}(P_\text{d}&[k]) :=\frac{P_\text{d}[k]}{\eta_\text{d}(P_\text{d}[k])} \label{cons3}\\
 P_\text{c}[k] &P_\text{d}[k] = 0  \label{cons6} \\
 \underline{E} \le &E[k] \le \overline{E} \label{cons4} \\
 0 \le &P_\text{c}[k], P_\text{d}[k] \le \overline{P}  \label{cons5} 
\end{align}
\end{subequations}
for $k=0,\hdots,K-1$.To avoid simultaneous charging and discharging while allowing for distinct charging and discharging efficiencies, a non-convex complementarity constraint is introduced in~\eqref{cons6}~\cite{almassalkhiTPWRS2015}. 
Next, we present a convex linear formulation of the NLP.


\subsection{Constant Efficiency Linear Formulation}
In this section, we present the convex linear formulation for the BESS model by assuming constant efficiencies equal to peak value of part-load efficiency. The nonlinear SoC estimation model given by \eqref{cons1} is relaxed to a linear model as follows,
\begin{equation}
   E[k+1] = E[k] + \frac{\Delta t}{C_{\text{batt}}} \left( \eta_\text{c} P_\text{c}[k] - \frac{1}{\eta_\text{d}} P_\text{d}[k] \right). \label{linear-cons1} 
\end{equation}

To achieve a convex formulation while satisfying complementary constraint, a two-stage method is implemented to address the non-convexity of \eqref{cons6} \cite{mazenCDC2023}. In the first stage, a cutting plane constraint \cite{boyd2007localization} is added to the formulation which reduces the effect of simultaneous charging and discharging and is given by,
\begin{equation}
    P_\text{c}[k] + P_\text{d}[k] \leq \overline{P}  \label{relax}
\end{equation}
This relaxed model permits simultaneous charging and discharging. To address this, a threshold $\alpha$ is applied to the net power $P_\text{net}[k] = P_\text{c}[k]- P_\text{d}[k]$ in the second stage, which selectively enforces complementarity. If $P_\text{net}[k] \geq \alpha$, discharging power is set to zero ($P_\text{d}[k]= 0$), and if $P_\text{net}[k] \leq -\alpha$ charging power is constrained to zero ($P_\text{c}[k]= 0$). The problem is then re-solved with the complementarity enforced at the appropriate timesteps, improving physical realizability while maintaining computational efficiency. While this approach can cause sub-optimality, it guarantees that the overall optimization problem remains convex~\cite{Elea}. In the next section, we introduce the ICNN formulation, which considers the nonlinear behavior of efficiencies.


\subsection{Relaxed ICNN formulation}
 \noindent In the previous sections, we presented the linear and nonlinear ERM formulation. To overcome the non-convexity of the nonlinear ERM, it can be approximated using an ICNN. The ICNN model begets

\begin{subequations}
{\small
\begin{align}
E[k+1] = E[k] + \frac{\Delta t}{C_{\text{batt}}}\left(f_{\text{ICNN}}(P_\text{c}[k]) + g_{\text{ICNN}}(P_\text{d}[k])\right)
\end{align}}
\end{subequations}
\noindent while $f_{\text{ICNN}}(P_\text{c}[k])$ and $g_{\text{ICNN}}(P_\text{d}[k])$, mentioned in \eqref{f-ICNN}-\eqref{g-ICNN}  ,are convex functions, but they do not form convex constraints due to the nonlinearity introduced by the point-wise maximum function. To handle this nonlinearity, a convex relaxation method is implemented. The relaxation for $z_{\text{c},i}[k] = \text{max}(0,W_{\text{c},i} z_{\text{c},i-1}[k]+D_{\text{c},i} P_\text{c}[k]+b_{\text{c},i})$ at layer $i$ is given by, 
\begin{subequations}
 {
 \begin{align}
z_{\text{c},i} [k] &\geq 0 \label{RelaxedICNN-1} \\
 z_{\text{c},i} [k] &\geq W_{\text{c},i} z_{\text{c},i-1}[k] + D_{\text{c},i} P_\text{c}[k] + b_{\text{c},i}. \label{RelaxedICNN-2} 
 \end{align}}
\end{subequations}
 \noindent The \eqref{RelaxedICNN-1} and \eqref{RelaxedICNN-2}  ensure that $z_{\text{c},i}[k]$ is always greater than or equal to both 0 and $ W_{\text{c},i} z_{\text{c},i-1}[k]+D_{\text{c},i} P_\text{c}[k] +b_{\text{c},i}$. This force $z_{\text{c},i}[k]$ to take the maximum of these two values. However, this relaxation only ensures that $z_{\text{c},i}[k]$ is greater than or equal to the maximum term, not necessarily equal to it. This can lead to suboptimal solutions, where $z_{\text{c},i}[k]$ overestimates the true value of the max function. To address this issue and tighten the relaxation, a penalty term is introduced into the objective function to minimize $z_{\text{c},i}[k]$. By penalizing $z_{\text{c},i}[k]$, we encourage the optimization to select the smallest possible values of $z_{\text{c},i}[k]$ that still satisfy the constraints \eqref{RelaxedICNN-1} and \eqref{RelaxedICNN-2}. This effectively forces $z_{\text{c},i}[k]$ to take the maximum of the two terms, thereby reducing the conservatism introduced by the relaxation. Similarly, an auxiliary variable $z_{\text{d},i}[k]$ is also introduced for the discharging term $g_{\text{ICNN}}(P_\text{d}[k])$. 
Now, the relaxed optimization problem is given as,

\begin{subequations}
\begin{align}
&\underset{P_\text{c}[k], P_\text{d}[k],z_{\text{c},i}[k],z_{\text{d},i}[k]}{\text{minimize}} \quad  f_0(P_\text{c}[k], P_\text{d}[k])\, +  \notag \qquad  \\
  \quad \quad &\phantom{MMMMM}\lambda \sum_{k=0}^{K-1} \sum_{i=1}^{N} (z_{\text{c},i}[k]+z_{\text{d},i}[k]) \\
\text{subject to} & \notag \\
E[k+1] &= E[k] + \frac{\Delta t}{C_{\text{batt}}}(f_{\text{ICNN}}(P_\text{c}[k]) + g_{\text{ICNN}}(P_\text{d}[k]))\\
& z_{\text{c},i} [k] \geq 0 \label{Relaxed-ICNN1} \\
& z_{\text{c},i} [k] \geq W_{\text{c},i} z_{\text{c},i-1}[k] + D_{\text{c},i} P_\text{c}[k] + b_{\text{c},i} \label{Relaxed-ICNN2} \\
& z_{\text{d},i} [k] \geq 0 \label{1} \\
& z_{\text{d},i} [k] \geq W_{\text{d},i} z_{\text{d},i-1}[k] + D_{\text{d},i} P_\text{d}[k] + b_{\text{d},1} \label{Relaxed-ICNN3} \\
& P_\text{c}[k] + P_\text{d}[k]\leq 1  \label{Relaxed-ICNN4}\\
& \underline{E} \leq E[k+1] \leq \overline{E} \label{Relaxed-ICNN5} \\
& 0 \leq P_\text{c}[k], P_\text{d}[k] \leq \overline{P}  \label{Relaxed-ICNN6} 
\end{align}
\end{subequations}
for $k=0,\hdots, K-1$ and $i=1,\hdots,N$ with $z_{\text{c},\text{N}}[k]=f_{\text{ICNN}}(P_\text{c}[k]) $ and $z_{\text{d},\text{N}}[k]=g_{\text{ICNN}}(P_\text{d}[k]) $.  The $\lambda$ is a penalty coefficient that balances the trade-off between the original objective function $f_0(P_\text{c}[k], P_\text{d}[k])$ and the minimization of the relaxed terms, $z_{\text{c},i}[k]$ and $z_{\text{d},i}[k]$. In this section, we introduced a relaxed model for the ICNN. To achieve a tighter relaxation, a Big-M formulation method can be employed, although this approach introduces binary integers into the formulation.

\subsection{Big-M ICNN formulation}

In this section, a Big-M formulation method is implemented to simplify the non-convex constraints related to $f_{\text{ICNN}}(P_\text{c}[k])$ and $g_{\text{ICNN}}(P_\text{d}[k])$. This formulation is widely used in mixed-integer programming (MIP) to handle disjunctions, i.e., cases where one of several conditions is active~\cite{Bazaraa}. The Big-M formulation introduces auxiliary variables and relaxed constraints, which effectively replace the max function in a way that preserves convexity while ensuring the correctness of transformation. The Big-M formulation for $z_{\text{c},i}[k] = \text{max}(0,W_{\text{c},i} z_{\text{c},i-1}[k]+D_{\text{c},i} P_\text{c}[k]+b_{\text{c},i})$ is given by,

\begin{subequations}
{\small
\begin{align}
F_{\text{c},i} [k] &\geq 0 \label{bigM-1} \\
F_{\text{c},i}[k] &\geq W_{\text{c},i} z_{\text{c},i-1}[k] + D_{\text{c},i} P_\text{c}[k] + b_{\text{c},i} \label{bigM-2} \\
F_{\text{c},i} [k] &\leq M(1 - y_i[k]) \label{bigM-3} \\
F_{\text{c},i} [k] &\leq W_{\text{c}_i} z_{\text{c},i-1}[k] + D_{\text{c},i} P_\text{c}[k] + b_{\text{c},i} + M y_i[k],
\label{bigM-4}
 \end{align}}
 \end{subequations}
\noindent where $F_{\text{c},i}[k]$ is an auxiliary variable that will take the value of the $\text{max}(0,W_{\text{c},i} z_{\text{c},i-1}[k]+D_{\text{c},i} P_\text{c}[k]+b_{\text{c},i})$ term, $y_i[k] \in \{0,1\}$ is a binary decision variable, and the Big-M constant $M$ is a sufficiently large positive value that ensures the constraint works as expected without cutting off the feasible solutions. \eqref{bigM-1} and \eqref{bigM-2}  ensure that $F_{\text{c},i}[k]$ is always greater than or equal to both 0 and $ W_{\text{c},i} z_{\text{c},i-1}[k]+D_{\text{c},i} P_\text{c}[k] +b_{\text{c},i}$. This force $F_{\text{c},i}[k]$ to take the maximum of these two values, which effectively models the max function. \eqref{bigM-3} and \eqref{bigM-4} control whether $F_{\text{c},i}[k]$ takes the value 0 or $ W_{\text{c},i} z_{\text{c},i-1}[k]+D_{\text{c},i} P_\text{c}[k]+b_{\text{c},i}$. When $y_i[k]=0$, \eqref{bigM-3} becomes inactive due to the large value of M, and \eqref{bigM-4} enforces $F_{\text{c},i}[k]= W_{\text{c},i} z_{\text{c},i-1}[k]+D_{\text{c},i} P_\text{c}[k]+b_{\text{c},i}$, which corresponds to the non-zero case of the max function. When $y_i[k]=1$, \eqref{bigM-3} forces $F_{\text{c},i}[k]=0$, corresponding to case when max equals zero.\\
The constant $M$ must be chosen carefully. It should be large enough to ensure the binary variable $y_i[k]$ properly switches between the two cases ($0$ and $W_{\text{c},i} z_{\text{c},i-1}[k]+D_{\text{c},i} P_\text{c}[k] +b_{\text{c},i}$). However, if $M$ is too large, it can introduce numerical instability or slow convergence in the optimization algorithm. A too small value does not guarantee the convergence to the same optimum of the original problem. Thus, $M$ should reflect the feasible range of the variables involved. The Big-M formulation is applied similarly to the discharging ICNN model. The MIP formulation is also implemented for the complementary constraint,
\begin{subequations}
    \begin{align}
        0 &\leq P_\text{c}[k] \leq w[k] \overline{P}\\
        0 &\leq P_\text{d}[k] \leq (1-w[k]) \overline{P}\\
        w&[k] \in {0,1}
    \end{align}
\end{subequations}
for $k = 0, \hdots,K-1$ and where the binary variable is such that if $w[k]=1 \Rightarrow P_\text{d}[k] = 0, P_\text{c}[k] \in [0 , \overline{P}]$. This method effectively preventing charging and discharging at the same time. While Big-M formulations can simplify problem structures, they introduce nonlinearity, making the overall problem more computationally challenging to solve.


\section{Power Systems Battery Use-Case: Revenue Maximization}
\label{use_cases}
In this section, we present a power system use-case of BESS to compare different BESS model formulations in terms of optimality, computational complexity, and practical implementability. Specifically, we focus on a revenue maximization scenario, where the BESS interacts with the grid under a given price signal to maximize overall economic returns. The objective of revenue maximization problem is to optimally dispatch a standalone battery in order to maximize the revenue generated from discharging energy to the grid, while simultaneously minimizing the costs associated with charging energy from the grid, given an energy price signal. The objective function for this problem can be written as:

\begin{flalign}
\underset{P_\text{c}, P_\text{c}}{\text{max}} \quad & \sum_{\text{k=0}}^\text{K-1}  \text{LMP}[k](P_\text{d}[k]- P_\text{c}[k]) \Delta t 
\label{rev_max}
\end{flalign}
where $\Delta t$ is the time step and LMP is locational marginal price.\\
The solution of the optimization problem is considered feasible for implementation if it can be applied to the BESS plant without violating energy or power constraints. 
 When this solution is implemented, and the SoC reaches its limits, the BESS control logic enforces the constraints by setting  $P_\text{c}[k]=0$ at the upper bound $\overline{E}$ and $P_\text{d}[k]=0$ at the lower bound $\underline{E}$. This enforcement can introduce discrepancies between the predicted and actual SoC trajectories, potentially leading to realized performance that is worse than the predicted performance. Such deviations can negatively affect long-term battery revenue, SOH, and grid reliability. Therefore, in this study, the solution generated by the optimization problem is referred to as the predicted solution, while the solution after implementation in the BESS plant is termed the actual solution. The predicted and actual revenues of each BESS model are analyzed to assess the optimality and feasibility of the solutions. \\
\begin{remark}[Integration of ICNN BESS Model in Networked Frameworks]
The ICNN BESS model interfaces with the grid only through the charging and discharging power variables, ($P_\text{c}[k]$ and $P_\text{d}[k]$). This design enables its integration into complex network frameworks, such as OPF problems. In this study, LMPs are treated explicitly, but in OPF problems, they appear implicitly as dual variables linked to power balance constraints.  While this implicit treatment does not increase computational complexity related to ICNNs, it can introduce challenges in analytical interpretation. 
\end{remark}


\subsection{Implementation}

The code base for the revenue maximization problem is developed using JuMP (a mathematical optimization package in Julia). This code base includes four models as discussed in Section \ref{methods}: (i) an NLP model, (ii) a linear model, (iii) a relaxed ICNN model, and (iv) a Big-M ICNN model. The nonlinear problem was solved using the Interior Point Optimizer (IPOPT), while the optimization problems for both the ICNN and linear models were implemented using the Gurobi optimizer. The BESS parameters used in this simulation study are given in Table~\ref{Spec}. The performance of these models is compared based on their predicted and actual objective values, as well as solver time.

\begin{table}[!htb]
\centering
\caption{BESS Parameters}
\label{Spec}
\begin{tabular}{>{\raggedright\arraybackslash}m{0.62\columnwidth} >{\raggedright\arraybackslash}m{0.12\columnwidth} >{\raggedright\arraybackslash}m{0.05\columnwidth}}
\toprule
\textbf{Parameters} & \textbf{Value} & \textbf{Unit} \\
\midrule
Max charge/discharge efficiency, $\eta_{\text{c,max}}/\eta_{\text{d,max}}$ & 0.92/0.95 & \quad - \\
Charge/discharge power rating, $\overline{P}$ & 50.0 & kW \\
Energy capacity rating, ${C_{\text{batt}}}$ & 135.0 & kWh \\
Maximum energy capacity, $\overline{E}$ & 0.9 ${C_{\text{batt}}}$ & kWh \\
Minimum energy capacity, ${\underline{E}}$ & 0.1 ${C_{\text{batt}}}$ & kWh \\
\bottomrule
\end{tabular}
\end{table}

\begin{figure}[t]
\centering
\includegraphics[width=0.45\textwidth]{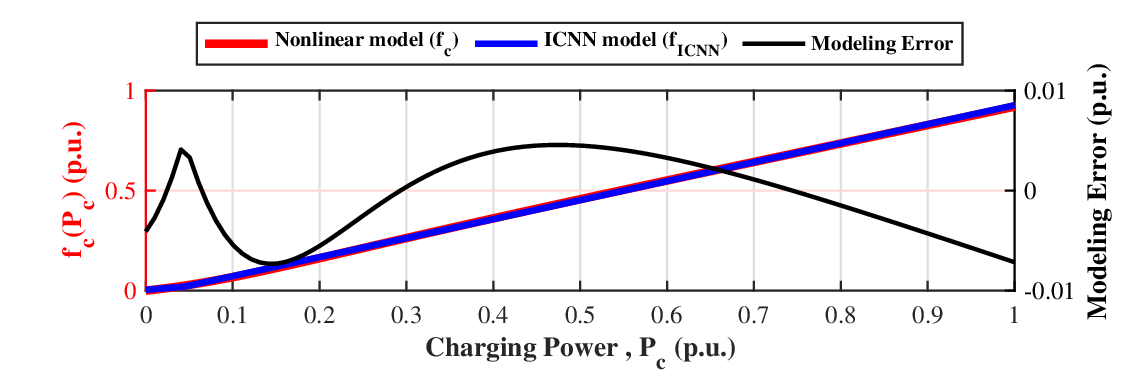}
\caption{ICNN-based approximation of charging $f_{\text{ICNN}}(.)$ }
\label{model}
\centering
\end{figure}

The data required for training of the ICNN consist of DC and AC power for the fully charging and discharging processes, which involves only a few hours of potential charging and discharging experiments. The plant model used to generate data is based on the nonlinear efficiency function presented in~\cite{Tanmay}. The Adam optimizer is employed to train the ICNN~\cite{Adam}. The trained neural network utilized in our work is a two-layer ICNN. This architecture was selected to balance the complexity of the model and the computational efficiency. The Fig.~\ref{model} presents the ICNN approximation of the nonlinear ERM for the charging process, represented by $f_{\text{ICNN}}(P_c)$. It is observed that in the low power region ($P_c\in[0.08,0.25]$ p.u.), the ICNN underestimates the nonlinear model. Conversely, for $P_c\in[0.25,0.8]$ p.u., the ICNN overestimates the model. For $P_\text{c}\in[0.05,1]$,  the modeling error does not exceed $10\%$. This modeling error may lead to SoC saturation.

\subsection{Results}
\begin{figure}[!tbp]
    \centering
    \begin{subfigure}[b]{1\columnwidth}
        \centering
        \includegraphics[width=0.98\textwidth]{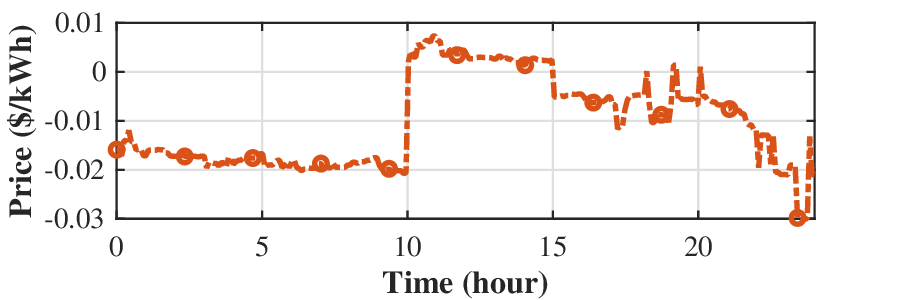} 
        \caption{}
        \label{LMP}
    \end{subfigure}
    \hfill
    \begin{subfigure}[b]{1\columnwidth}
        \centering
        \includegraphics[width=0.98\textwidth]{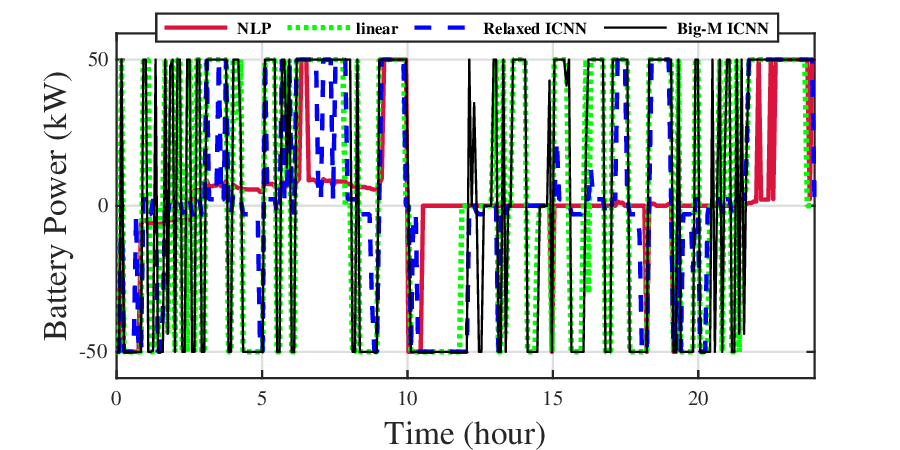} 
        \caption{} 
        \label{Rev_max_p}
    \end{subfigure}
    \begin{subfigure}[b]{1\columnwidth}
        \centering
        \includegraphics[width=0.98\textwidth]{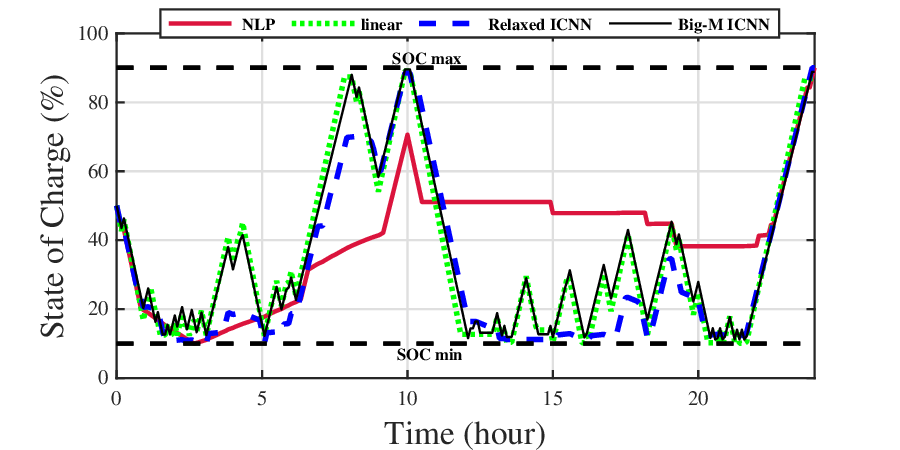} 
        \caption{} 
        \label{Rev_max_soc}
    \end{subfigure}
    \caption{Comparison of the simulation results for the revenue maximization problem using nonlinear, linear, Relaxed ICNN (with $\lambda = 4.3\times10^{-3}$), and Big-M ($M=4$) formulations (a) Price (LMP) (b) Optimal battery dispatch (c) Battery SoC}
    \label{Revenue_figures}
\end{figure}

This subsection presents the simulation results for the revenue maximization problem. Fig.~\ref{LMP} shows the 24-hour price signal at 5-minute intervals for this problem.
The battery power dispatch and SoC profiles are shown in Fig.~\ref{Rev_max_p} and \ref{Rev_max_soc}, respectively. A different battery dispatch from the linear model results in a difference in SoC trajectory. As a result, the linear model overestimates the system's capabilities, which results in errors in system performance and revenue projections.

The performance of all four methods is summarized in Table~\ref{Revenue Maximization}. The nonlinear model shows a suboptimal solution due to the nonconvexity of the problem, resulting in the worst solution. The linear ERM model predicts the highest revenue; however, its actual revenue upon implementation is lower than that of the relaxed and Big-M ICNN models due to its neglect of nonlinear efficiencies. Although the relaxed and Big-M ICNN models predict lower revenues compared to the linear model, their performances are better in terms of feasibility. The actual revenue of the Big-M ICNN model is higher than that of the relaxed ICNN model; however, this comes at the cost of increased computational time, which may present challenges for larger problems. The discrepancy between the actual error and the predicted error in the Big M ICNN model results from the modeling error.

\begin{table}[!htb]
\centering
\caption{Revenue Maximization Simulation Results}
\label{Revenue Maximization}
\begin{tabular}{>{\raggedright\arraybackslash}m{0.45\columnwidth} >{\raggedright\arraybackslash}m{0.1\columnwidth} >
{\raggedright\arraybackslash}m{0.12\columnwidth} >{\raggedright\arraybackslash}m{0.12\columnwidth}}
\toprule
\textbf{Model} & \textbf{Solver Time(s)} & \textbf{Predicted Revenue (\$)} & \textbf{Actual Revenue (\$)} \\
\midrule
Nonlinear ERM & 1.95 & 3.04 & 3.04 \\
Linear ERM & 0.013 & 5.36 & 4.75 \\
Relaxed ICNN ($\lambda = 4.3\times10^{-3}$) & 0.021 & 4.97 & 4.97 \\
Big-M ICNN ($M=4$)& 1.33 & 5.33 & 5.24 \\
\bottomrule
\end{tabular}
\end{table} 

\begin{figure}[h]
    \centering
    \includegraphics[width=\columnwidth]{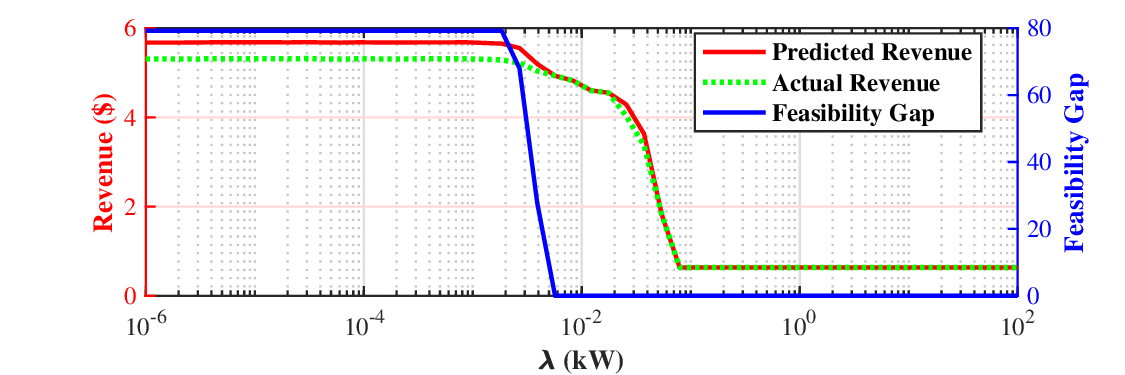} 
    \caption{Impact of penalty coefficient $\lambda$ on optimality and feasibility gaps (predicted and actual objective values, as well as the aggregated relaxation gap for the revenue maximization problem).}
    \label{lambda_b}
\end{figure}

\subsection{Discussion}
Fig.~\ref{lambda_b} shows the effect of $\lambda$ on both actual and predicted revenues, as well as the aggregated feasibility gap for the relaxed formulation of ICNN. When $\lambda$ is small, the feasibility gap is large, and there is a gap between actual and predicted revenues. As $\lambda$ increases, both the feasibility gap and revenues decrease. Which again indicates degraded performance. When $\lambda$ is zero, the revenue is highest, but this occurs at the expense of optimality. These results suggest a clear trade-off between performance and optimality, and $\lambda$ must be carefully chosen to balance the both.

\begin{figure}[t]
        \centering
        \includegraphics[width=0.8\columnwidth]{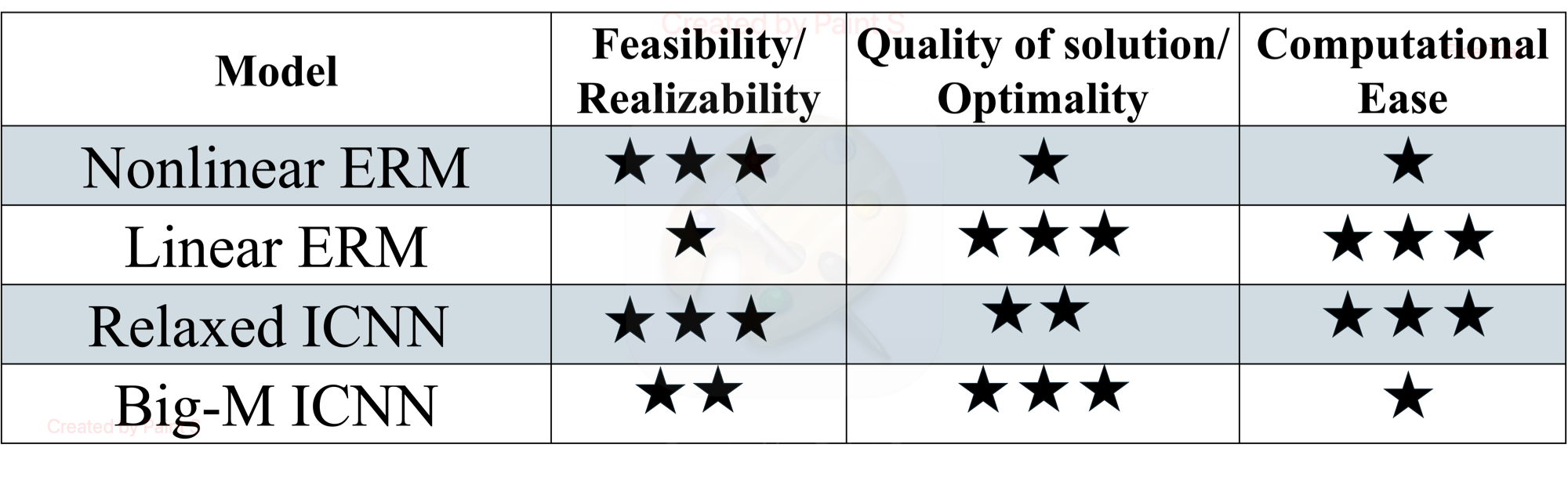}
        \caption{Qualitative comparison of BESS formulations}
        \label{Comparison}
        \centering
\end{figure}

The qualitative comparison of the four methodologies is summarized in Fig.~\ref{Comparison} which exhibit distinct trade offs in the optimization of BESS. The nonlinear ERM shows high accuracy and feasibility but is constrained by non-convexity, leading to suboptimal performance. Conversely, the linear ERM offers a lower computational burden and enhanced optimality; however, this comes at the expense of accuracy, potentially resulting in an overestimation of system capabilities. Even though Big-M ICNN model performances better than other three methods, but faces challenges associated due to computational burden due to MIP formulation. These findings suggest that the relaxed ICNN model offers a promising practical solution, particularly suited for real-time grid applications, where a balance between computational burden and model accuracy is critical. 

\section{CONCLUSIONS}
This paper introduces a new ICNN-based approach to model a nonlinear representation of the ERM, including part-load efficiencies. The ICNN trains a neural network with constrained weights and ReLu activation function to engender a convex approximation of the nonlinear part-load inverter efficiency curve. However, within an optimization formulation, the ICNN's equality constraint is non-convex. 
To address this challenge, we propose two approaches: (i) a relaxed convex ICNN-based formulation and (ii) a Big-M ICNN mixed-integer (MIP) formulation.

The performance of the relaxed ICNN methodology is compared with three other methods for a whole-sale market revenue maximization application. The relaxed ICNN method appears promising, as it strikes a balance between accuracy, optimality, and computational efficiency when carefully designed. Future work will advance KKT analysis to characterize conditions under which relaxed ICNNs yield a zero duality gap. We also consider extending data-driven methods, such as ICNNs, to higher-fidelity battery models.
\bibliographystyle{IEEEtran}
\bibliography{refs} 
\end{document}